# The effect of elastic-plastic mismatch and interface proximity on the fracture toughness of Ti-TiN thin films

Nidhin George Mathews[1,a], Aloshious Lambai[1], Marcus Hans[2], Jochen M. Schneider[2], Gaurav Mohanty[1], Balila Nagamani Jaya[3,a]

[1] *Materials Science and Environmental Engineering, Faculty of Engineering and Natural Sciences, Tampere University, Tampere 33014, Finland*

[2] *Materials Chemistry, RWTH Aachen University, 52074 Aachen, Germany*

[3] *Department of Metallurgical Engineering and Materials Science, Indian Institute of Technology Bombay, Mumbai 400076, India*

[a] *Corresponding authors: nidhin.mathews@tuni.fi; jayabalila@iitb.ac.in*

## Abstract

Magnetron sputtered titanium nitride (TiN) thin films are widely used as protective coatings due to their high hardness, but suffer from inherent brittleness and low fracture toughness, limiting their applicability. The multilayering of TiN films with metallic titanium (Ti) interlayers in the form of bi-layer and tri-layer systems have been studied using microcantilever fracture tests. Plastic dissipation in the Ti layer is shown to lead to an increase in crack growth resistance. The effect of the elastic-plastic mismatch between the two materials on the crack driving force, as well as the size of the fully developed plastic zone in Ti have been quantified in this work for the first time. It is shown that incorporating a Ti layer thickness of 250 nm can improve the fracture resistance by nearly ten times compared to the initiation fracture toughness in TiN, preventing catastrophic fracture of these multi-layered films. These results will aid in physics informed design of optimised thickness of metallic interlayers in multi-layered thin films.

## 1 Introduction

Titanium nitride (TiN) provides excellent hardness, mechanical strength, wear resistance, corrosion resistance and chemical stability which find applications in harsh or extreme environmental conditions. It has been used extensively as a hard coating material for different applications in cutting tool manufacturing, energy applications, aerospace and automotive parts, and microelectromechanical devices [1–4]. These applications demand the coatings to sustain high mechanical stresses without compromising their structural integrity. Even though TiN possesses high compressive strengths, the material is inherently brittle and has low fracture toughness. The high degree of internal stresses and thermal mismatch makes it difficult to produce monolithic TiN films with good adhesion to the substrate beyond a certain thickness [5]. Combining the TiN coating with a metallic titanium (Ti) layer further has been shown to improve the durability of the multilayered architectures by enhancing its fracture resistance [6–11].

Metal-ceramic nanolaminates are shown to have high hardness and strength as the layer thickness is reduced. In extremely thin nanolaminates of metal-ceramic architectures, the ceramic layer co-deforms with the metallic layer due to crossing of dislocations [12,13]. The Ti-TiN multilayered architectures are fabricated in order to have composite mechanical properties of high hardness, strength and toughness. Tailoring the layer thickness of the Ti-TiN multilayer architecture changes the residual stress state, along with bringing an elastic-plastic mismatch across the interface, which in turn contributes to the improvement in mechanical properties such as hardness ($H$), elastic modulus ($E$), and fracture toughness ($K_{IC}$) of layered architectures [7,14–19]. In comparison to the monolithic TiN coatings, Ti-TiN multilayers exhibit good impact resistance as the Ti layer was shown to dissipate part of the energy through plastic deformation [19]. The increase in thickness of the Ti layer reduces the residual stresses in the Ti-TiN architecture [5] and improves the adhesion between the layers, but the increase in volume fraction of Ti layer lowers the overall hardness of the multilayers [20,21]. Tailoring the Ti layer thickness is therefore essential to using the multilayer in actual applications. The reduction in the period thickness of Ti/TiN film and therefore increased number of interfaces for the same volume fraction were shown to improves the hardness and wear resistance of the films [22,23]. Interestingly, the control of interface structure can induce plasticity in the ceramic layer, which in turn improves the wear resistance. The typical sharp interface in Ti-TiN was modified into gradient interfaces using nitrogen compositional variation, which lead to reduced wear rate than monolithic TiN [24].

Several previous studies have been carried out to determine the fracture behaviour of Ti-TiN multilayer architectures using indentation-based techniques [7,9,10,21]. They showed an improvement in the fracture resistance of the architecture with the increase in the layer thickness and thickness ratio of the Ti (ductile) layer with respect to the TiN (brittle) layer, consequently with a drop in hardness. But these tests do not provide any insights into the nature of the fracture, since the fracture path under a complex tri-axial stress state of an indenter is unknown. Notched microcantilever bending has been widely used to study the crack growth behaviour of free-standing and multilayered films, as one can determine the contribution of individual layers and interfaces towards crack propagation [25–27]. Fracture toughness ($K_{IC}$) values of sputtered TiN monolayers are reported to lie in the range of 1.2 – 3.0 MPa.m$^{0.5}$, which fail in a linear elastic brittle manner [28–33]. In a specific magnetron sputtered film, the addition of ductile Ti layer was shown to provide toughening, increasing the fracture toughness to 3.2 MPa.m$^{0.5}$ for a 50 layer Ti-TiN multilayer, almost double that of monolayer TiN films which showed a fracture toughness of 1.4 MPa.m$^{0.5}$ [11]. Crack growth under mode I loading conditions was found to be straight, cutting across the layers, following the inter-columnar boundaries of the sputtered films. The Ti/TiN interface was found to be strong and prevented any crack deflection along the interface. The elastic-plastic mismatch between the Ti and TiN layers was quantified in terms of changes in crack driving force, which was accounted for in their modified stress intensity factor solutions [11]. In addition, the compressive residual stresses (up to 8 GPa) which developed due to the layering may also have imparted additional toughening, which could not be separately discerned. However, while the sputtered Ti layer independently showed some degree of crack tip plasticity and non-linear deformation behaviour, it did not exhibit any expression of plasticity when sandwiched between TiN in the multilayer architectures. The possible reasons for this suppression of crack tip plasticity are due to the constraint on both sides of the Ti layer by the harder and stiffer TiN, and the inability to accommodate plasticity in individual Ti layer due to its low thickness (80 nm). However, there was uncertainty due to the low stiffness of the indentation-micromechanical loading system, which could have overloaded the microcantilevers, leading to abrupt, catastrophic fracture [11].

The objective of this study is to address these open questions on the effect of elastic-plastic mismatch between the layers and its relative position with respect to crack tip on promoting crack tip plasticity and prevention of catastrophic fracture in these architectures. Notched microcantilevers of trilayered Ti-TiN-Ti, and two bilayered Ti-TiN and TiN-Ti (micromachined out of the same Ti-TiN-Ti trilayer sputtered films) are the focus of the study. The crack tip position is varied to alternately determine the effect of an elastically and plastically stiffer (TiN) and a compliant (Ti) layer ahead of the interface, followed by the overall fracture behaviour of the trilayer architecture. The presence of a ductile, softer, metallic layer is expected to a) modify the crack driving force due to elastic-plastic shielding/anti-shielding at the interface, b) crack tip blunting and, c) crack closure due to compressive residual stresses. Finite element simulations are performed to estimate the crack driving force variation with the specific conditions of bilayer and trilayer architectures.

## 2  Materials and methods

### 2.1  *Thin film deposition and microstrucutral characterisation*

Multilayer thin film samples of Ti and TiN were synthesised on Si (100) substrate using high power pulsed magnetron sputtering in an industrial-scale CemeCon CC800/9 chamber. The trilayer film architecture of Ti-TiN-Ti was deposited with Ti (nominal thickness $W_3$ = 1.6 μm) as the first layer on Si, while the subsequent TiN and Ti layers were supposed to exhibit nominal thicknesses of $W_2$ =1.3 μm and $W_1$ = 1.1 μm respectively. Details about the deposition parameters are described in a previous work [11]. The strucutral characterisation of the trilayer film was performed using grazing incidence X-ray diffraction (GI-XRD) studies (Malvern Panalytical Emperyon, United Kingdom). XRD scans were performed on the film at an incidence angle of 10º using CuK$_\alpha$ radiation of wavelength (λ) 0.154 nm, operated at 40 kV voltage and 30 mA current. Focussed ion beam (FIB) cross-sectioning was performed on the Ti-TiN-Ti trilayer architecture to determine individual layer thicknesses. The Ga$^+$ ion current at 300 pA at an accelerating voltage of 30 kV was selected for the cross-section milling. The milled region was imaged in ion channeling contrast in a dual beam SEM (Zeiss Crossbeam 540, Germany) using the 50 pA ion current to visualise the layers. The hardness of the trilayer film was measured to be 8 GPa, using nanoidentation,in our previous study [11], whereas that of the monolithic Ti and TiN were measured to be 6.3 GPa and 23.7 GPa respepctively. The elastic modulus of the monolithic Ti and TiN layers were found to be 148.5 GPa and 377.7 GPa respectively, using nanoindentation [11].

### 2.2  *Microcantilever sample preparation and fracture testing*

Micromechanical fracture testing of the thin film samples was performed using single-edge notched microcantilever bend tests. Free-standing microcantilevers were fabricated at the edge of the specimen using FIB milling (Zeiss Crossbeam 540, Germany). Ion currents of 15 nA, 3 nA and 0.3 nA at 30 kV operating voltage were selected for the coarse, intermediate and fine milling steps respectively. The front and back sides of the cantilevers were fine polished at an additional tilt angle of ±1.5º to maintain uniform rectangular cross-section of the beam. The geometrical dimensions of the cantilever, length (*L*): thickness (*W*): width (B) were maintained at constant ratio of 4.5: 1: 1 for all cases. Notches were milled at a distance (*x/W*) ~ 2 μm from the fixed end of the cantilever using 10 pA milling current. Notch depth to width ratio (*a/W*) was maintained in the range 0.25 - 0.3 in all test cases with the notch positioned entirely in the top layer. These geometrical dimensions and the loading position were considered so as to maintain a dominant mode I loading conditions for fracture in the cantilevers [34]. Fig. 1a - Fig. 1c shows the images of microcantilevers of each film-substrate condition: a) Ti-TiN bilayer ($W_{1-2}$ = 2.4 µm) with the notch in the Ti layer, b) TiN-Ti bilayer ($W_{2-3}$ = 2.4 µm) with the notch in the TiN layer and c) Ti-TiN-Ti trilayer ($W_{1-3}$ = 3.7 µm) with the notch in the Ti layer. The layer which is mentioned first is on the top surface of the cantilever. It should be reiterated that the bilayer TiN-Ti and Ti-TiN cantilevers were prepared from the trilayer film by FIB milling the top and bottom layers respectively. In case of the crack lying in the Ti layer, or in close proximity to it, a plastic zone size ($r_y$) of ~300 nm was estimated using the Irwin process zone model, according to Eq. 1, considering $K_{IC}$ = 3 MPa.m$^{0.5}$ and $\sigma_y$ = 2.1 GPa, where $\sigma_y$ is the nominal yield strength in bending, derived from microcantilever bend tests [11]. The estimated plastic zone size is ~10 times lower than the cantilever dimensions *(B, W)* ensuring plane strain conditions of fracture.

$$r_y = \frac{1}{2\pi}\left(\frac{K_{IC}}{\sigma_y}\right)^2 \qquad\qquad Eq.\ 1$$

The microcantilever fracture tests were performed using an *in situ* displacement controlled nanoindenter (Alemnis AG, Switzerland) inside a SEM (Zeiss Leo 1450, Germany). A diamond wedge indenter tip (Synton MDP, Switzerland) of 10 µm length and 100º opening angle was used for loading the cantilevers, to ensure line contact and prevent indentation based deformation [35]. Fracture experiments were conducted in displacement control mode with tip travel speed of 50 nm/s. In such displacement-controlled tests, crack propagation results in distinct load drops. The displacement values recorded during the tests were then corrected for the machine compliance to get the actual displacement of the cantilever. At least three microcantilevers were tested for each condition. Initiation fracture toughness ($K_{IC}$) of the tested films based on linear elastic fracture mechanics (LEFM) was calculated from the critical load ($P_c$) *i.e.* maximum load at which crack initiation occurs, using Eq. 2 [25] and Eq. 3 [34], assuming linear elastic behaviour until then. Note: The value of *W* in the subsequent equations can take the corresponding values of $W_{1-2}$, $W_{2-3}$ and $W_{1-3}$ depending on the architecture selected.

$$K_{IC} = \frac{P_c L}{B W^{1.5}} f\left(\frac{a}{W}\right) \qquad\qquad Eq.\ 2$$

$$f\left(\frac{a}{W}\right) = -3.15 + 72.85\left(\frac{a}{W}\right) - 188.51\left(\frac{a}{W}\right)^2 + 202.61\left(\frac{a}{W}\right)^3 \qquad Eq.\ 3$$

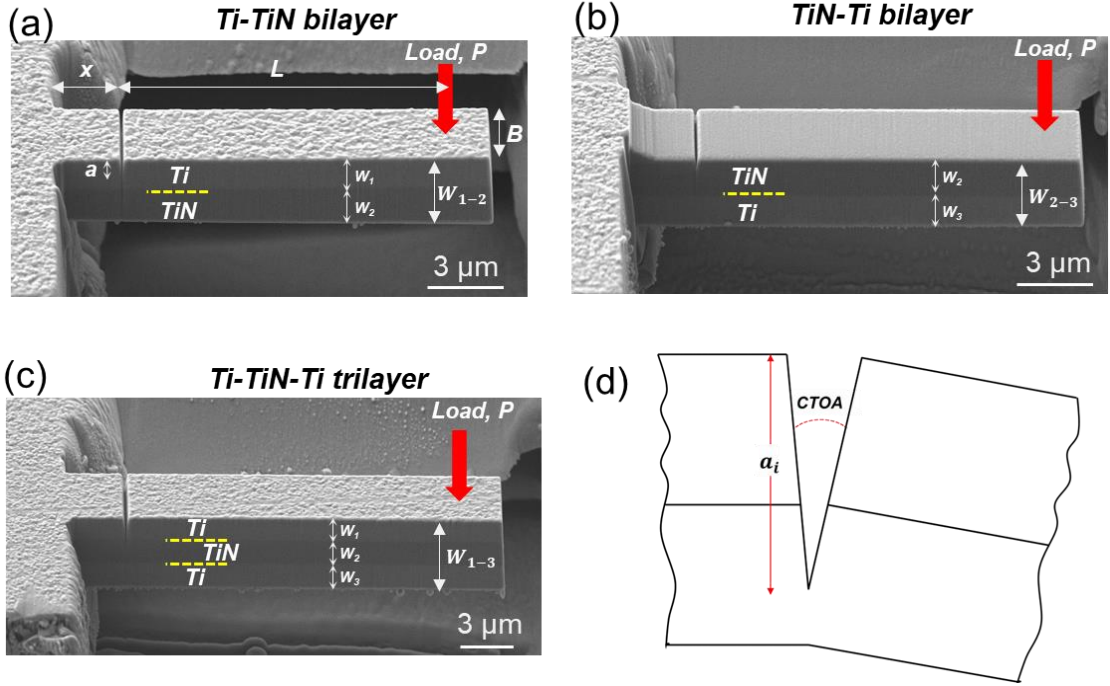

*Fig. 1: Single edged notched microcantilever samples of a) Ti-TiN bilayer, b) TiN-Ti bilayer, and c) Ti-TiN-Ti trilayer architecture used for fracture testing. (Note: The initial notch was always positioned within the top layer). d) Schematic showing the measurement of crack tip opening angle for crack length 'a' at instant 'i'.*

To quantify the fracture behaviour in the non-linear elastic-plastic regime of fracture (*i.e.* post the first load drop), crack tip opening angle (CTOA) was used as the fracture resistance parameter of material. CTOA is defined as the included angle between two faces of the opening crack surfaces measured with crack tip as the vertex (as shown in Fig. 1d) [36]. The overall fracture resistance ($J_R$) is also evaluated considering the total area under load-displacement curve using Eq. 4 [37], according to elastic plastic fracture mechanics (EPFM).

$$J_R^{(i)} = \frac{\eta\, A_{tot,i}}{B\,(W-a_i)} \qquad Eq.\ 4$$

where, $A_{tot,i}$ is the total area under the load-displacement curve at instant *i* (or more specifically the total work done during the event), $\eta$ is a shape factor equal to 2 for cantilevers, and $a_i$ denotes the crack length at instant *i*.

## 2.3 Finite element simulations procedure

The procedure to determine $K_{IC}$ according to LEFM assumes a homogeneous material system, which is not strictly the case here. Hence, the crack driving force at the crack tip ($J_{tip}$) developed during the microcantilever fracture experiments were estimated from finite element simulations using a two-dimensional model in ABAQUS CAE (Dassault Systemes Simulia Corp.). The material properties of the layers were assumed to be isotropic and homogeneous within the layers, with TiN ($E$ = 377.7 GPa, $\sigma_y$ = 7.9 GPa [11]), and Ti ($E$ = 148.5 GPa, $\sigma_y$ = 2.1 GPa [11]). Strain hardening in the Ti layer was incorporated using the Ramberg-Osgood plasticity model [38]. The interface was assumed to be sharp and adherent. The material properties used for Ti and TiN layers in the model have been reported in a previous study [11]. Boundary conditions of the cantilever were chosen such that the beams were encastered at one end without translations or rotations, while the other end was free to move in vertical direction. A wedge-shaped indenter was used to apply loads to the cantilever end. Surface-to-surface interaction was maintained with the outer surface of the indenter tip and the top surface of cantilever to ensure proper contact while loading. A stationary (non-propagating) crack was defined as a seam inside the cantilever. The simulations were performed for the following cases: i) Ti-TiN bilayer, ii) TiN-Ti bilayer, iii) Ti-TiN-Ti trilayer, as shown in Fig. 2a – Fig. 2c. The notch depth was maintained such that two crack lengths were explored for each case, wherein the crack driving force at initiation, and the shielding/anti-shielding effects as the crack approaches the interface are both estimated. For the bilayer cantilevers, $a/W$ of 0.3 (crack away from the interface) and 0.5 (crack near the interface) was selected, whereas these ratios were 0.3 and 0.65 for the trilayer cantilever. The interface is nearly at half of the total film thickness for bilayer whereas the interface is at 2/3$^{rd}$ of the thickness for the trilayer case. The crack driving forces were estimated using the contour integral approach with refined meshes near the crack tip (Fig. 2d) as recommended by Treml et al. [39].

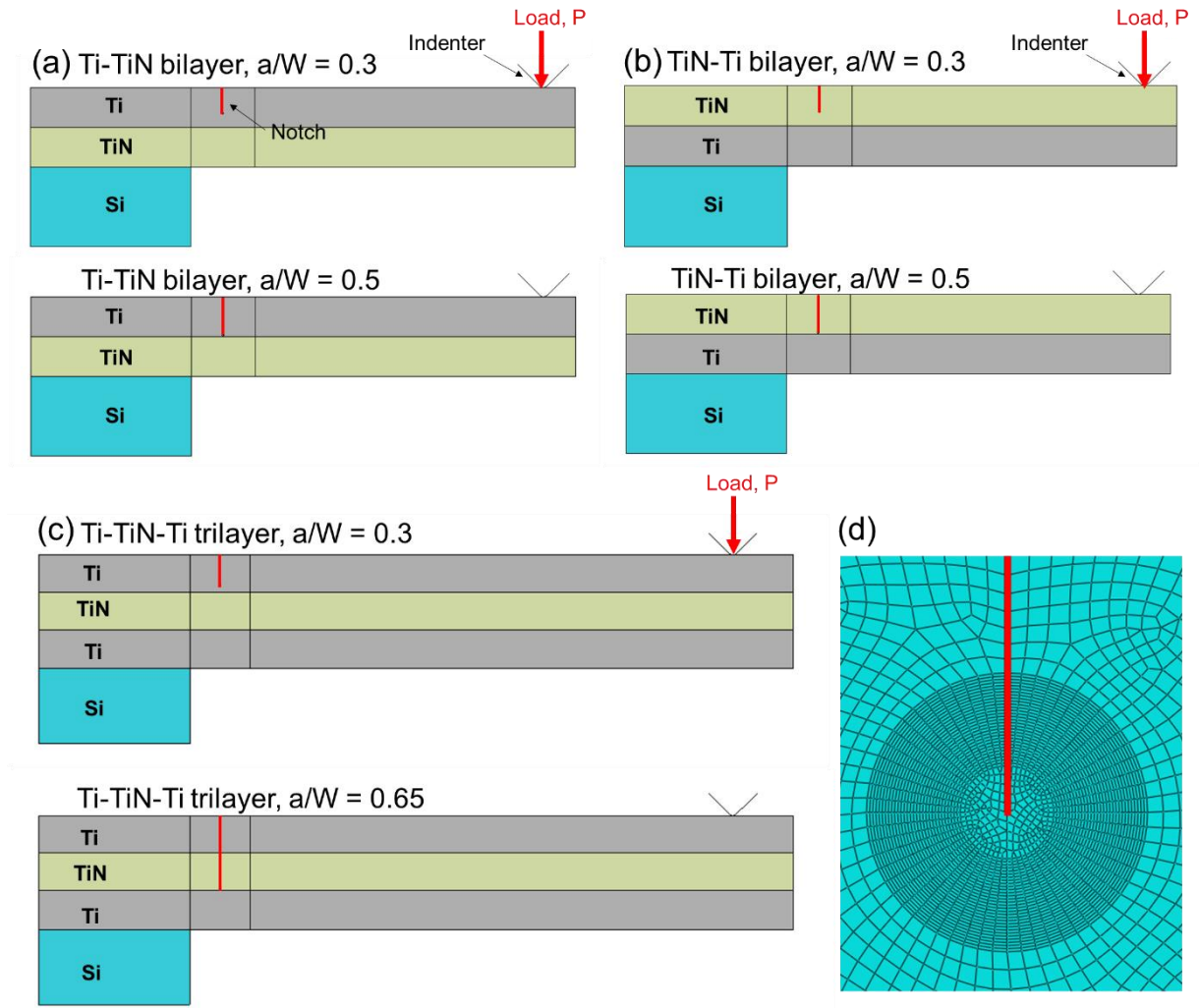

*Fig. 2: Finite element model of a) Ti-TiN bilayer, b) TiN-Ti bilayer, and c) Ti-TiN-Ti trilayer architecture used for simulations with corresponding a/W ratios. d) Finite element mesh using near the crack tip to estimate the crack driving force.*

## 3 Results

### 3.1 *Microstructure of Ti-TiN thin film architectures*

The GI-XRD pattern of the trilayer Ti-TiN-Ti thin film sample is shown in Fig. 3a. The XRD pattern confirms the hexagonal phase of Ti and cubic phase of TiN from corresponding peak positions, and shows the polycrystalline nature of the layers. The peaks of TiN are lower in intensity and exhibit a larger full width at half maximum due to the smaller grains compared to the Ti layers. The ion channeling contrast image of the film cross-section as shown in Fig. 3b reveals the columnar growth pattern of the films and its nanocrystalline grain structure. A sharp and well adhered interface is visible with each individual layer of thickess of ~1.6 µm (Ti layer on Si substrate), ~1.3 µm (TiN layer on Ti) and ~1.1 µm (Ti layer on TiN). Crystallite sizes of approximately 29 nm for Ti and 5 nm for TiN films were estimated using the Scherrer formula [40], from the corresponding full width at half maximum (FWHM) of the (002) peak.

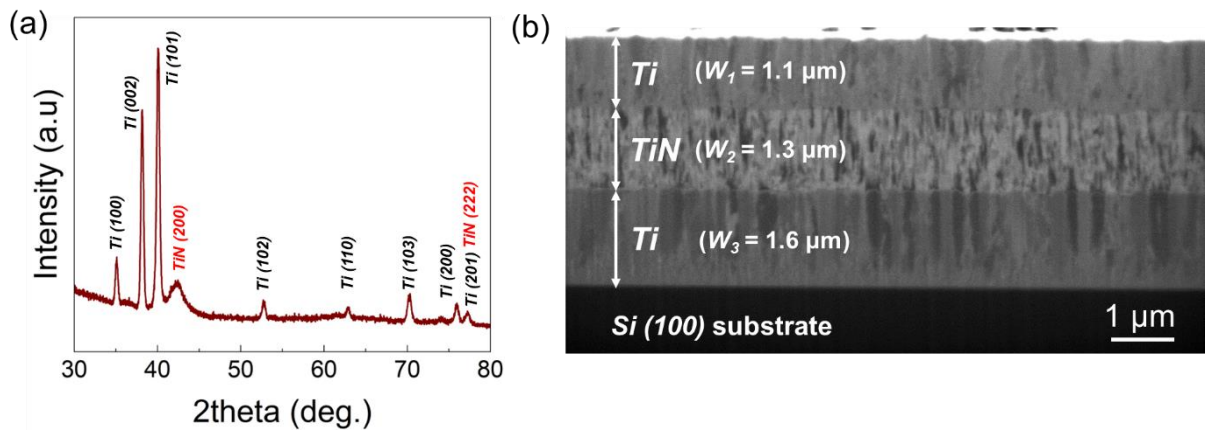

*Fig. 3: a) Grazing incidence XRD pattern showing the phases of Ti (hexagonal) and TiN (cubic), and polycrystalline nature of films. b) Ion channelling image showing grain contrast of Ti-TiN-Ti trilayer film architecture.*

### 3.2 Fracture behaviour of bilayer and trilayer Ti-TiN architectures

The representative load-displacement curves and corresponding post-fracture images of the Ti-TiN bilayer, TiN-Ti bilayer and Ti-TiN-Ti trilayer microcantilever fracture tests are shown in Fig. 4. All bilayer and trilayer architectures showed an initial linear elastic response followed by a sharp load drop which corresponds to the crack propagation at the initial notch (Fig. 4a). The peak load was found to be the highest for the Ti-TiN-Ti trilayer, while for Ti-TiN and TiN-Ti bilayers the peak loads reached to less than 50% of the trilayer peak load. In the case of bilayer Ti-TiN, a complete brittle fracture of the whole system occurred as the crack propagated catastrophically across the brittle TiN layer without any crack arrest (Fig. 4b). In contrast to this, for the bilayer TiN-Ti architecture, the crack growth was arrested at TiN/Ti interface as the crack encountered a ductile Ti layer ahead (Fig. 4c). The load-displacement curve did not show a load drop to zero, unlike in the Ti-TiN case. Instead, a partial load shedding was followed by an increasingly non-linear deformation of the cantilever, due to the plastic zone expansion in the Ti layer. Both bilayer architectures showed a similar slope in the initial elastic loading part as the combined stiffness remains the same in both configurations. This is different in case of Ti-TiN-Ti trilayer configuration, which showed a higher slope. Considering the overall width of the specimen being larger in this case, this is expected. Post the large load drop at the first peak load, during which the crack propagates across the Ti/TiN interface and past the TiN layer to reach the next TiN/Ti interface, the load rises again due to the plastic dissipation in the bottom Ti layer (Fig. 4d).

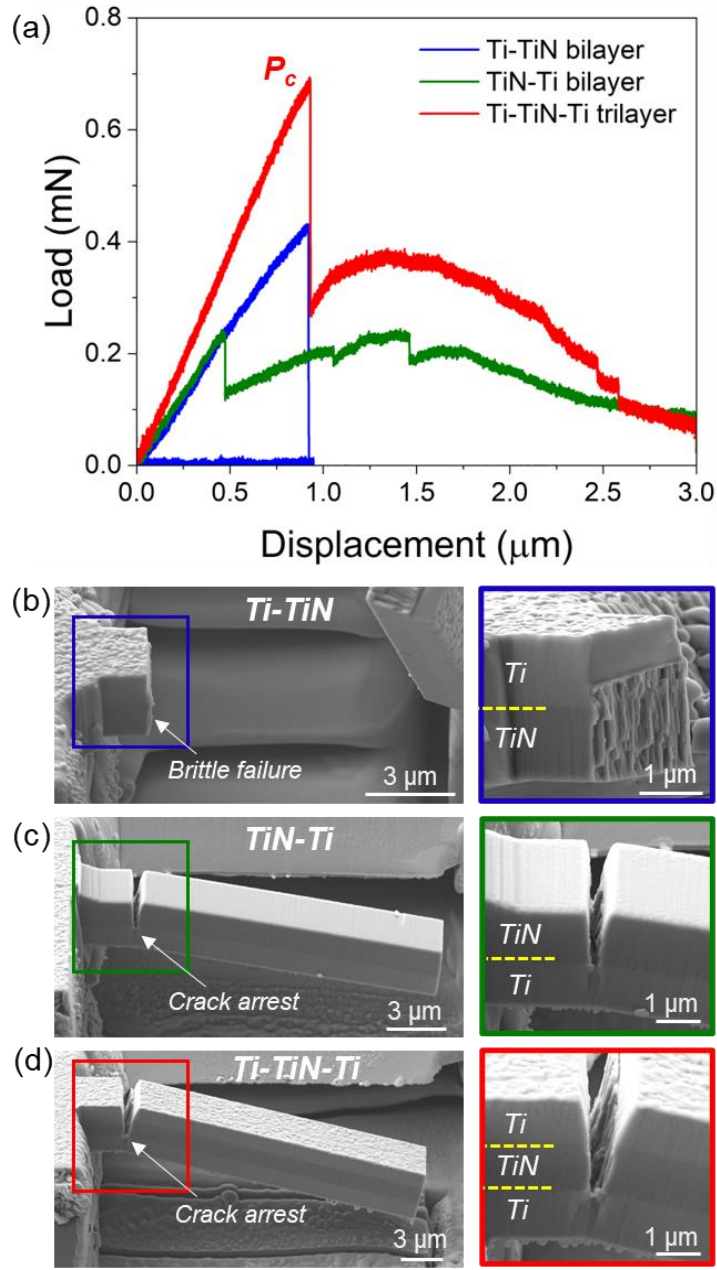

*Fig. 4: a) Representative load-displacement curves showing the fracture behaviour of different thin film architectures during the microcantilever fracture experiments. Post fracture images of the microcantilevers showing b) brittle failure in Ti-TiN bilayer, crack arrest in c) TiN-Ti bilayer, and d) Ti-TiN-Ti trilayer architecture.*

To summarise, in the Ti-TiN bilayer the crack propagates catastrophically in to the TiN layer in a linear elastic manner leading to complete fracture of the beam, whereas in the TiN-Ti and Ti-TiN-Ti architectures, the crack propagates rapidly up to the TiN/Ti interface during the first load drop, post which the plastic zone in the Ti layer prevents it from further propagation, leading to gradual crack growth. Since the initial crack growth occurs post linear elastic deformation, LEFM criterion can be used to determine the fracture toughness. The initiation $K_{IC}$ obtained in terms of LEFM based calculations using Eq. 2 and Eq. 3, for each architecture is listed in Table 1. Average $K_{IC}$ of 2.7 ± 0.7 MPa.m$^{0.5}$ for Ti-TiN, 1.6 ± 0.4 MPa.m$^{0.5}$ for TiN-Ti, and 3.2 ± 1.0 MPa.m$^{0.5}$ for Ti-TiN-Ti was estimated. The initiation $K_{IC}$ values of Ti-TiN and Ti-TiN-Ti architectures are comparable to the reported $K_{IC}$ values of the individual Ti film (3.0 ± 0.4 MPa.m$^{0.5}$) [11], as the crack lay in the Ti layer in these cases, while that of the TiN-Ti architecture was closer to the individual TiN film (1.5 – 2.0 MPa.m$^{0.5}$) [41,30,11,42]. Other values of $K_{IC}$ have also been reported for TiN (between 2.3 – 2.8 MPa.m$^{0.5}$) [28–33,42–44]. The possible causes of variation in the $K_{IC}$ value can be the difference in processing parameters, internal microstructure of the film, thickness of the film and in some cases, type of fracture testing method used [45,46]. However, for the sputtered film tested here, the fracture toughness of free-standing TiN tested using the same microcantilever bend technique was found to be 1.52 ± 0.4 MPa.m$^{0.5}$ [11].

For the bi-layer TiN-Ti and the tri-layer Ti-TiN-Ti, the pronounced plasticity in the Ti layer after initial crack growth forces an elastic-plastic fracture mechanics based evaluation of fracture resistance using the J-integral approach. The iterative method of determining the plasticity contribution of J-integral solution using the unloading stiffness for each crack increment (as mentioned in [37]) is not used here as the system is not homogeneous, and shows abrupt, discrete load drops (Fig. 4a). The fracture resistance was determined using Eq. 4 considering the area until the first load drop for initiation toughness ($J_c$) and the total area for overall fracture resistance ($J_R$), and is listed in Table 1. The initiation $J_c$ values for Ti-TiN and TiN-Ti architectures are found to be 44 N/m and 23 N/m respectively. The initiation $J_c$ for the trilayer is determined to be 52 N/m. This $J_c$ follows the same trend as a $K_{IC}$, which is as expected given the linear elastic regime of deformation at this stage. Further crack growth was quantified in terms of fracture resistance $J_R$. The Ti-TiN does not show any crack growth resistance at all, with catastrophic fracture ensuing post initiation. The pronounced plasticity at the crack tip near the Ti layer improves the $J_R$ in TiN-Ti and Ti-TiN-Ti, which were estimated to be 353 ± 3 N/m and 498 ± 33 N/m respectively. Corresponding CTOA was found to be 18º and 19º respectively, indicating that $J_R$ is a better measure to differentiate the fracture resistance between the two architectures than CTOA. This work is the first to show the contribution of plasticity in Ti to the overall fracture resistance as well as the first such attempt to quantify the fracture resistance of Ti-TiN multilayers using the elastic-plastic fracture mechanics parameters. This approach can be utilized in the future to estimate the fracture resistance of multilayers with optimized spacing and volume fraction.

*Table 1: Summary of the fracture resistance parameters obtained for each film architecture from microcantilever fracture tests.*

| Film architecture | Average initial crack length (μm) | Fracture toughness, $K_{IC}$ from LEFM (MPa.m$^{0.5}$) | $J_c$ from EPFM (N/m) | $J_R$ from EPFM (N/m) | Crack tip opening angle, CTOA (º) |
|---|---|---|---|---|---|
| Ti-TiN with crack in Ti | 0.8 | 2.7 ± 0.7 | 44 ± 6 | 44 ± 6 | - |
| TiN-Ti with crack in TiN | 0.6 | 1.6 ± 0.4 | 23 ± 1 | 353 ± 3 | 18 ± 2.0 |
| Ti-TiN-Ti with crack in Ti | 1.0 | 3.2 ± 1.0 | 52 ± 16 | 498 ± 33 | 19 ± 0.1 |

# 4  Discussion

Fig. 5a is a schematic representing the fracture mechanism in the three cases. In the Ti-TiN case, the distance between the initial notch and the Ti/TiN interface is too small for a well developed plastic zone to form in the Ti layer. This truncation of the plastic zone leads to an almost brittle, catastrophic fracture of the same. The post-fracture images of the Ti-TiN cantilever surface shown in Fig. 4b confirms the brittle fracture, as the crack propagated along the intercolumnar boundaries of the films of both Ti and TiN layers without any hinderance. For the TiN-Ti case, as soon as the crack reaches the interface, the plasticity in the Ti layer prevents further crack growth and the expansion of the process zone is not hindered, except upon reaching the free surface at the bottom (Fig 5a). For Ti-TiN-Ti try-layer as well, although the plastic zone in the top Ti layer is truncated due to the TiN layer ahead of it, the bottom Ti layer is free to deform plastically and expand its process zone. Therefore, for the TiN-Ti and Ti-TiN-Ti architectures, the plastic zone formation in the Ti layer led to crack arrest and toughening, preventing complete fracture of the beam (Fig. 4c and 4d).  There is clear evidence of such a plastic deformation zone and generation of nanocracks in the Ti layer in both architectures (Fig. 5b and 5c). This shows that plasticity in the Ti layer can indeed be leveraged to improve the fracture resistance of multi-layered TiN-Ti films.

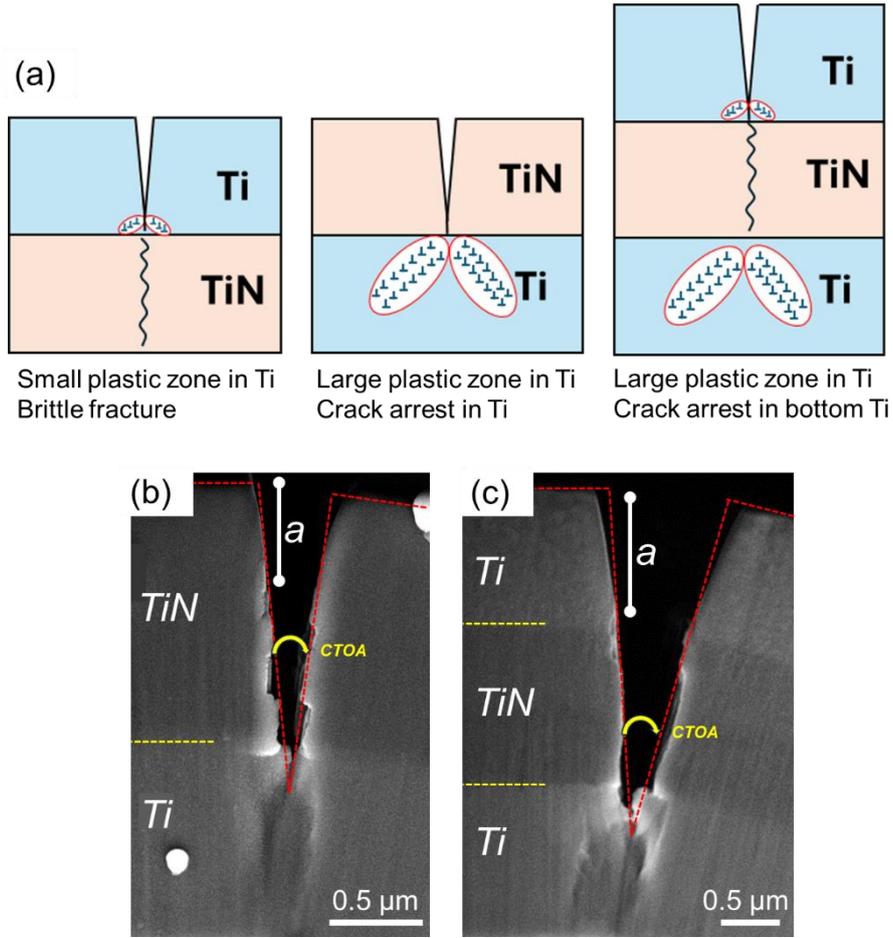

*Fig. 5: a) Schematic showing the plastic zone development in Ti layer and the crack growth and crack arrest in each layer. Crack arrest and plastic zone development in Ti layers in b) Ti-TiN and c) Ti-TiN-Ti architectures at maximum displacement of the cantilever (with crack tip opening angle marked in red lines).*

    Both high hardness and fracture resistance is necessary for hard protective coatings for their improved performance. While hardness variations have been well documented in the literature as a function of layer thickness and volume fraction, factors that control fracture resistance in metal-ceramic multilayers have not yet been clarified in the literature. The current study focusses on the effect of crack tip plasticity on the fracture behaviour of the multilayer architecture relative to the positon of notch tip and its proximity to the interface. It also explores whether the plastic mismatch controls the fracture resistance compared to the elastic mismatch in the case of Ti-TiN. The elastic modulus difference between the high stiffness TiN and low stiffness Ti introduces elastic mismatch $\left(\frac{E_{TiN}}{E_{Ti}} \sim 2.5\right)$ and modifies the crack driving force due to shielding/anti-shielding effect as the crack approaches the respective interfaces [47,48]. The extent of elastic shielding is expected to be 12 % in terms of the crack tip driving force ($K_{tip}$) [49]. The hardness difference between the TiN and Ti introduces a plastic mismatch $\left(\frac{H_{TiN}}{H_{Ti}} \sim 4\right)$

as well. When the crack approaches the interlayer from the plastically stronger material, the crack tip driving force is amplified as the plastic zone spreads across the interlayer [50,51], leading to an anti-shielding effect. The extent of this effect depends on plastic zone size and the proximity of the crack to the interface, which governs the relative expansion of the process zone in the plastically soft layer. A rigid layer ahead of the crack constrains the plastic zone and truncates it, while a soft layer would allow for expansion of the plastic zone. In the present study, there is both elastic and plastic mismatch between Ti and TiN, leading to a reinforcement of shielding or anti-shielding effects. In order to quantify this, the J-integral calculations were carried out to determine the deviation in crack tip driving force ($J_{tip}$) from the $J_R$.

The $J_{tip}$ curve as a function of the cantilever displacement is shown in Fig. 6a. For a fixed displacement of 1 μm, the TiN-Ti has a higher $J_{tip}$ compared to the Ti-TiN architecture, while the trilayer Ti-TiN-Ti has the lowest crack driving force. Specifically, the longer crack in the TiN-Ti experiences a higher crack driving force than the shorter crack, while the longer crack in the Ti-TiN experiences a lower crack driving force than the shorter crack. This confirms the anti-shielding effect (an attractive force towards the crack due to the elastically and plastically compliant Ti layer) in the TiN-Ti system, and a shielding effect (a repulsive force towards the crack due to the elastcially and plastically stiff TiN layer) in the Ti-TiN system. The trilayer experiencing the lowest crack driving force amongst all has more to do with the larger specimen dimensions reducing the overall crack driving force for the same displacement. Such changes in crack driving force as the crack approaches an elastically stiffer or compliant layer ahead of it has been well documented in the literature [48,52]. However, in the present study, it is seen that the fracture resistance of the Ti-TiN architecture (2.7 ± 0.7 MPa.m$^{0.5}$) with the crack in the Ti layer is still very similar to the fracture resistance of the monolithic Ti layer (3.0 ± 0.4 MPa.m$^{0.5}$) [11]. Similarly, the fracture resistance of the TiN-Ti architecture (1.6 ± 0.4 MPa.m$^{0.5}$) with the crack in the TiN layer is nearly the same as the fracture resistance of monolithic TiN (1.5 ± 0.2 MPa.m$^{0.5}$) [11]. This is very different from the results obtained in bilayer BaTiO$_3$/SrTiO$_3$ as well as bilayer BaTiO$_3$/Pt-Si systems, where proximity to the interface resulted in a discernible increase in the fracture resistance of the system in the presence of a stiffer layer [52]. Since no such effect of the second layer is seen on the initiation fracture toughness of the Ti-TiN and TiN-Ti systems, the origin of the fracture resistance in these systems is further looked into from the point of view of plasticity in the Ti layer. Only one such study on the effect of yielding and hardening of a metallic layer on the fracture resistance of metal-ceramic films has been reported to the best of our knowledge on Cu-Si [53], and even these are based on numerical solutions.

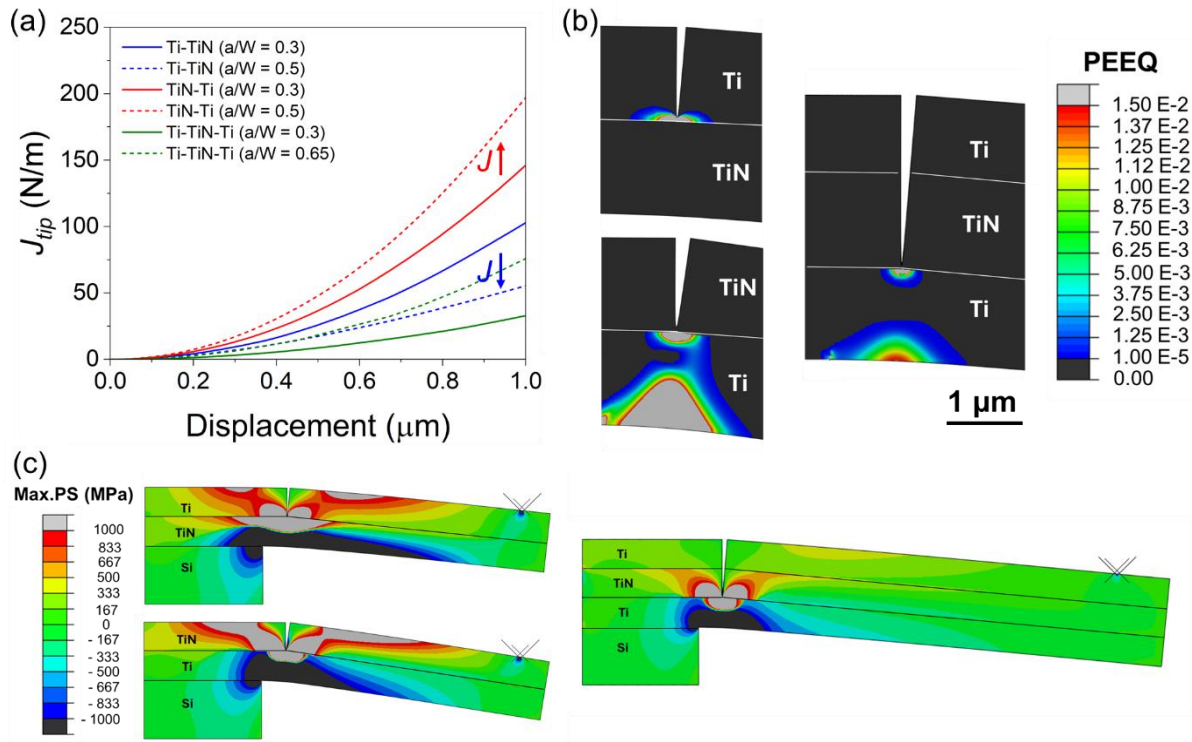

*Fig. 6: a) Crack driving force as a function of cantilever displacement in Ti-TiN, TiN-Ti and Ti-TiN-Ti architectures from finite element simulations assuming stationary, non-propagating cracks. b) Plastic zone evolution near the crack tip (in red circles) inside the Ti layer (shown as plastic strain equivalent - PEEQ) when the crack meets the TiN/Ti interface at the respective critical displacements, and c) maximum principal stresses (Max.PS) contours in Ti-TiN bilayer, TiN-Ti bilayer and Ti-TiN-Ti trilayer architecture at the respective critical displacements.*

In both bilayer configurations, the crack once initiated, extends rapidly towards the second layer until it encounters the interface, upon which the fracture behaviour diverges for the two cases. Further crack propagation depends on the ductile or brittle nature of the next layer ahead. The *R*-curve in the bilayer TiN-Ti and the trilayer Ti-TiN-Ti is made possible due to a fully developed plastic zone in the Ti layer once the crack reaches the TiN/Ti interface, whereas in case of Ti-TiN system, no plastic zone is formed in the TiN layer when the crack reaches the interface (Fig. 6b). Plastic zone formation in TiN is nearly impossible given that TiN is almost 4 times as strong as Ti. The plastic zone size is found to be of the order of 256 nm in the TiN-Ti bilayer and 277 nm in the Ti-TiN-Ti trilayer Fig. 6b, measured at their respective critical displacements, when the crack reaches that interface. This represents a fully developed plastic zone since there the layer thickness is at least 5 times the plastic zone size, with the free-surface not causing any truncation. In addition, the plasticity due to the compressive stresses developing on the bottom side of the beam (Fig. 6c), leads to additional toughening and preventing further crack growth in the bilayer TiN-Ti and the trilayer Ti-TiN-Ti (Fig. 6b). Such direct evidence of crack tip plasticity and large scale yielding in metallic Ti thin films has been reported for the first time in this study. When the Ti layer is reduced in thickness so as to truncate its plastic zone to less than 250 nm, such an *R*-curve behavior is not observed anymore [11]. In fact in such cases, the Ti film is expected to behave like a brittle material, rather than as a ductile metal. When the plastic zone at the crack tip is not allowed to expand due to the constraint in size or an interface with a hard layer ahead of it, it is uncommon to see plasticity in metallic materials, which drastically changes the fracture behaviour [54,55].

To summarise, on comparing the fracture resistance parameters of bilayer Ti-TiN and TiN-Ti architectures (Table 1), the crack located in the Ti layer (for Ti-TiN) shows higher initiation resistance compared to the case of crack in TiN layer (for TiN-Ti). This is expected given that the $K_{IC}$ of single layer Ti is almost double that of the TiN layer [11]. In comparison to single layer Ti, $K_{IC}$ of Ti-TiN is 10% lower, reinforcing that the elastic shielding effect of the TiN layer ahead of the crack tip plays an insignificant role, while truncation of crack tip plasticity in fact reduces the initiation $K_{IC}$ of the bilayer. In fact a stiffer layer resists the expansion of the plastic zone in the Ti layer from fully reaching its stable size. On the other hand, the $K_{IC}$ of TiN-Ti is ~7% higher than the single layer TiN, since a softer Ti ahead of the crack tip allows for relaxing the stress intensification in the TiN. The $K_{IC}$ of the trilayer 16% higher than that of the monolithic Ti, since the the truncation of the plastic zone by the TiN is countered by the crack tip plasticity in the third Ti layer. In case of the trilayer, the initial crack experiences a shielding effect due to the TiN layer ahead of it, and as the crack propagates to the TiN/Ti interface, it experiences toughening due to the crack tip plasticity in the Ti layer. Therefore, the trilayer is able to synergistically blend the advantages of elastic and plastic shielding to give the highest toughening. This is schematically represented already in Fig. 5a. The quantification of the *R*-curve in such base architectures of two layer and three layer systems will help in the design of multilayers and graded architectures with improved fracture resistance, without resorting to trial and error. Graded multilayers with a harder surface and a tougher interior can be designed based on this study, so as to leverage the advantage of resisting crack growth as fracture initiates from the surface and proceeds towards the interior.

# 5 Conclusions

The fracture behaviour of bilayered and trilayered Ti-TiN multilayered film architectures were studied using microcantilever fracture tests to understand the fundamental crack growth mechanics and mechanisms in metal-ceramic based thin films. The initiation fracture toughness values of 2.7 ± 0.7 MPa.m$^{0.5}$ and 1.6 ± 0.4 MPa.m$^{0.5}$ were found for Ti-TiN and TiN-Ti layers, similar to $K_{IC}$ of monolithic Ti and TiN respectively. More importantly, it was shown that including a Ti layer ahead of the crack tip leads to a rising fracture resistance, with the plastic dissipation in the Ti layer preventing catastrophic fracture, in TiN-Ti bilayer and Ti-TiN-Ti trilayer architectures, unlike in the case of Ti-TiN bilayer. These results show that the position of the crack tip and its proximity to the interface, as well as the type of layer ahead of it determines the fracture behaviour of multi-layered thin films. In elastic-bilayers such as ceramic/ceramic multilayers, the shielding effect of the stiffer layer improves the fracture resistance by imposing crack closure forces at the crack tip. Unlike this, these results show that the elastic shielding ahead of the crack tip plays an antagonistic role by suppressing crack tip plasticity, and hence reducing the initiation toughness of metal-ceramic multilayers. On the other hand, plastic dissipation in the layer ahead of the crack tip improves propagation resistance by expanding the process zone. The critical Ti thickness needed for plastic dissipation to aid in improved fracture resistance is found to be nearly 250 nm. These insights will aid in physics informed design of Ti-TiN multilayers with optimum thickness of the metal interlayer which prevents catastrophic fracture, while maintaining other properties such as high hardness, by leveraging the effect of elastic-plastic mismatch between the metal and the ceramic.

## CRediT authorship contribution statement

**Nidhin George Mathews**: Conceptualization, Methodology, Data curation, Formal Analysis, Investigation, Visualization, Writing – original draft, Writing – review & editing. **Aloshious Lambai**: Methodology. **Marcus Hans**: Methodology, Writing – review & editing. **Jochen M. Schneider**: Methodology, Writing – review & editing. **Gaurav Mohanty:** Supervision, Funding Acquisition, Writing – original draft, Writing – review & editing. **Balila Nagamani Jaya**: Conceptualization, Supervision, Writing – original draft, Writing – review & editing.

## Declaration of competing interest

The authors declare that they have no known competing financial interests or personal relationships that could have appeared to influence the work reported in this paper.

## Acknowledgements

Authors acknowledge the Research Council of Finland (grant no. 341050) for financial support. The work made use of facilities at Tampere Microscopy Centre at Tampere University.